\pdfoutput=1
\documentclass[11pt]{article}
\usepackage[margin=1in]{geometry}
\usepackage[utf8]{inputenc}
\usepackage[T1]{fontenc}
\usepackage{amsmath,amssymb,mathtools}
\usepackage{amsthm}
\usepackage{graphicx}
\usepackage{booktabs}
\usepackage[ruled,vlined]{algorithm2e}
\SetKwInput{KwIn}{Input}
\SetKwInput{KwOut}{Output}
\SetKw{Continue}{continue}
\DontPrintSemicolon
\usepackage{microtype}
\usepackage{tikz}
\usetikzlibrary{calc}
\usepackage{xcolor}
\usepackage[colorlinks=true,linkcolor=blue!60!black,citecolor=blue!60!black,urlcolor=blue!60!black,pdfauthor={Max Klimm},pdftitle={A Sublogarithmic Competitive Algorithm for Online Graph Exploration}]{hyperref}
\usepackage{thm-restate}
\usepackage{needspace}
\theoremstyle{plain}
\newtheorem{theorem}{Theorem}
\newtheorem{lemma}[theorem]{Lemma}
\newtheorem{corollary}[theorem]{Corollary}
\theoremstyle{definition}
\newtheorem{definition}[theorem]{Definition}
\renewenvironment{proof}[1][\proofname]{\par\medskip\noindent\textit{#1.}\ \ignorespaces}{\par\medskip}
\renewcommand{\qed}{\ifmmode\square\else\unskip\nobreak\hfill\ensuremath{\square}\fi}
\providecommand{\doi}[1]{\href{https://doi.org/\detokenize{#1}}{\url{https://doi.org/#1}}}
\providecommand{\mathsc}[1]{\text{\normalfont\scshape\MakeLowercase{#1}}}
\newcommand{\ALG}{\mathsc{ALG}}
\newcommand{\OPT}{\mathsc{OPT}}

\newcommand{\DFS}{\textnormal{DFS}}
\newcommand{\hDFS}{\textnormal{hDFS}}%
\newcommand{\hDFSs}{\hDFS^{*}}%
\newcommand{\Blocking}[1][\delta]{\textnormal{Blocking}_{#1}}
\newcommand{\Process}{\mbox{\textnormal{\textsc{Process}}}}
\newcommand{\Explore}{\mbox{\textnormal{\textsc{Explore}}}}
\newcommand{\R}{\mathbb{R}}

\DeclareMathOperator{\dist}{d}
\newcommand{\MST}{\mathsc{MST}}
\newcommand{\MSF}{\mathsc{MSF}}
\newcommand{\hDFSB}{\hDFSs_{b}}
\newcommand{\hDFSBu}{\hDFSs_{b,u}}
\newcommand{\eps}{\varepsilon}
\newcommand{\E}{\mathbb{E}}

\title{A Sublogarithmic Competitive Algorithm\\for Online Graph Exploration}
\author{Max Klimm\\[2pt]
{\small Institute of Mathematics, TU Berlin}\\
{\small Department of Economics, Sciences Po}}
\date{}

\begin{document}
\pagestyle{plain}
\maketitle
\thispagestyle{plain}
\begin{abstract}
In online graph exploration, an agent must visit every vertex of an unknown weighted graph and return to its start, learning the incident edges at each vertex it visits.
On $n$-vertex graphs, the best known competitive ratio has been $O(\log n)$, achieved by several algorithms, among them the hierarchical depth-first search of Megow, Mehlhorn, and Schweitzer.
We give an algorithm with competitive ratio ${(2+o(1))}\ln n/\ln\ln n$ for known~$n$.
Like hierarchical depth-first search, it groups the edges into geometric weight classes and searches depth-first between components of lighter edges.
Unlike it, the agent tours each component along a minimum spanning tree for the original weights, so only crossings between components pay for rounding, and the base $b$ of the weight class can grow with~$n$.
The leading constant $2$ is tight for our choice of base.
Without knowing $n$, restarts with increasing size guesses give upper bounds with leading constant $4$ for a fixed schedule and $2$ in expectation for a randomly shifted schedule.
If the weights lie in a known range $[1,2^k]$ with $k\ge4$, choosing $b=k/\log_2 k$ gives a competitive ratio of $O(k/\log k)$ without knowing $n$.
\end{abstract}

\section{Introduction}\label{sec:intro}

Online graph exploration models an agent mapping an unknown environment while paying for every move.
Starting at a vertex $s$ of an unknown connected weighted graph, the agent learns, at each vertex $v$ it visits, the identifier of every neighbour $u$ and the weight of the edge $\{u,v\}$. It must visit every vertex and return to $s$.
The cost is compared with $\OPT$, the length of a shortest closed walk visiting all vertices.
This local-information version of the travelling salesman problem was introduced by Kalyanasundaram and Pruhs~\cite{kalyanasundaram1994}.

Every visit to a new vertex takes the agent across a \emph{boundary edge}, an edge with one visited and one unvisited endpoint. To reach the next unvisited vertex, the agent pays both for travel to the visited endpoint and for crossing the edge itself. An expensive edge may be available at the agent's current vertex, while a cheaper edge may lie at the far end of the visited graph.
Three algorithms resolve this trade-off in different ways and achieve a competitive ratio of $O(\log n)$ on general $n$-vertex graphs. For nearest neighbour, the logarithmic bound follows from the analysis of Rosenkrantz, Stearns, and Lewis~\cite{rosenkrantz1977} applied to the shortest-path metric of the graph. The agent minimizes the total distance to the next unvisited vertex. It can therefore take a local edge arbitrarily heavier than a distant one if reaching the distant edge costs enough. The power-of-two version of hierarchical depth-first search ($\hDFS$) of Megow, Mehlhorn, and Schweitzer~\cite{megow2012} gives priority to lighter weight classes. It can cross an arbitrarily long visited path to reach an edge of weight $1$ before taking a local edge of weight $2$.

Blocking makes the distance it is willing to travel depend on the weight of the local edge~\cite{kalyanasundaram1994,megow2012}. It postpones a boundary edge $e$ if the unvisited endpoint of a strictly lighter boundary edge is reachable from the visited endpoint of $e$ by a path of length at most $(1+\delta)w(e)$ whose internal vertices are already visited. Baligács et al.~\cite{baligacs2023jcss} show that choosing $\delta=\log n$ gives Blocking a competitive ratio of $O(\log n)$ on general graphs.
In 1994, Kalyanasundaram and Pruhs~\cite{kalyanasundaram1994} asked whether a constant competitive ratio is achievable. Even whether any algorithm achieves $o(\log n)$ has remained open~\cite{brandt2020}.

\subsection{Results and techniques}\label{sec:results}

We give a deterministic algorithm for general $n$-vertex graphs with a competitive ratio of $O(\log n/\log\log n)$, whether or not the agent knows $n$. This improves the classical $O(\log n)$ bounds for nearest neighbour~\cite{rosenkrantz1977} and power-of-two $\hDFS$~\cite{megow2012}, and the more recent $O(\log n)$ bound for Blocking with $\delta=\log n$~\cite{baligacs2023jcss}. At each weight threshold, keeping only edges of weight at most that threshold gives connected components called clusters. Our algorithm, $\hDFSB$, retains the hierarchy of Megow et al.\ and changes how the agent travels inside a fully visited cluster. Accounting for this travel at its original cost allows us to use wider weight classes. Table~\ref{tab:ratios} summarizes the resulting guarantees.
\begin{table}[tb]
\caption{Asymptotic upper bounds on the competitive ratio for $n$-vertex graphs. Randomized bounds hold in expectation against an oblivious adversary. In the last row, all edge weights lie in $[1,2^k]$.}\label{tab:ratios}
\centering
\small
\begin{tabular*}{\textwidth}{@{}l@{\extracolsep{\fill}\hspace{.5em}}r@{\extracolsep{0pt}\hspace{.3em}}l@{\extracolsep{\fill}\hspace{.5em}}r@{\extracolsep{0pt}\hspace{.3em}}l@{}}
\toprule
Setting & \multicolumn{2}{c}{Deterministic} & \multicolumn{2}{c@{}}{Randomized} \\
\midrule
Known $n$ & $(2+o(1))\dfrac{\ln n}{\ln\ln n}$ & [Thm.~\ref{thm:main}] & $(2+o(1))\dfrac{\ln n}{\ln\ln n}$ & [Thm.~\ref{thm:offset}] \\[6pt]
Unknown $n$ & $(4+o(1))\dfrac{\ln n}{\ln\ln n}$ & [Thm.~\ref{thm:phases}(b)] & $(2+o(1))\dfrac{\ln n}{\ln\ln n}$ & [Thm.~\ref{thm:phases}(a)] \\[6pt]
Unknown $n$, known $k$ & $O(k/\log k)$ & [Thm.~\ref{thm:range}] & \multicolumn{2}{c@{}}{---} \\
\bottomrule
\end{tabular*}
\end{table}
Within a fully explored component, the hDFS of Megow et al.~\cite{megow2012} visits the boundary edges in the depth-first order of a local minimum spanning tree and moves between them along shortest paths. For general weights, the whole algorithm runs on the graph with weights rounded up to powers of two, so that the local trees and the shortest paths all use rounded weights. We use the rounding, now to an arbitrary base $b$, only to define the classes. Inside a cluster, the agent follows a closed DFS tour of a minimum spanning tree for the original weights, so an internal tree edge costs twice its original weight each time its cluster is toured. Only crossings between clusters pay for rounding. This is what allows $b$ to grow with $n$.

The total cost of crossings is at most $2b\MST$, where $\MST$ is the weight of a minimum spanning tree of the graph. Summing the internal travel charges over tree edges and levels gives a bound of $2(1+\log_b(n-1))\MST$. An edge $e$ of a minimum spanning tree $T$ is charged at most once per level, and only at levels above its own class and up to the class of the heaviest edge of $T$. If this edge has weight $w_{\max}$, then $e$ is charged at most $1+\log_b(w_{\max}/w(e))$ times. Light edges can therefore be charged at many levels, but they carry little weight. By concavity of the logarithm, $\sum_{e\in T}w(e)\log_b(w_{\max}/w(e))\le\MST\cdot\log_b(n-1)$. Increasing $b$ thus raises the crossing bound and lowers the travel bound. For known $n$, choosing $b=\max\{2,\ln n/(\ln\ln n)^2\}$ makes the crossing term lower order and gives the leading constant $2$ in Theorem~\ref{thm:main}. A uniform random shift of the class boundaries sharpens the finite bound by averaging the rounding loss and the number of levels that charge each edge. The leading constant is again $2$ (Theorem~\ref{thm:offset}).

For unknown $n$, we restart deterministic exploration with increasing guesses for the number of vertices. With a fixed schedule, the bound on failed guesses has the same leading term as the bound on the successful phase, giving a leading constant of $4$ in the upper bound. Shifting the schedule by a single uniform random offset makes the expected cost of the failed phases lower order and recovers the leading constant $2$ (Theorem~\ref{thm:phases}).

A known weight range gives another way to choose the base without knowing~$n$. If all weights lie in $[1,2^k]$, taking $b=k/\log_2 k$ gives a competitive ratio of $O(k/\log k)$ as $k\to\infty$ (Theorem~\ref{thm:range}).
This improves, by a logarithmic factor, the $O(k)$ bound for power-of-two $\hDFS$ that follows from~\cite[Theorem~4]{megow2012}. The path in the proof of~\cite[Theorem~5]{megow2012} shows that this bound is tight for that algorithm.

For known~$n$, the analysis is tight (Theorem~\ref{thm:lower}, proved in Appendix~\ref{sec:tight}). For every choice of $b=b(n)\ge2$, $\hDFSB$ has a competitive ratio of $\Omega(\log n/\log\log n)$. For the stated base, the leading constant~$2$ is attained with and without the random offset.

\subsection{Related work}\label{sec:related}

Rosenkrantz et al.~\cite{rosenkrantz1977} give the upper bound $\frac12\lceil\log_2n\rceil+\frac12$ for nearest neighbour. Its logarithmic order is tight~\cite{hougardy2014,hurkens2004}.
Megow, Mehlhorn, and Schweitzer~\cite{megow2012} introduced $\hDFS$, proved a $2k$ bound for $k$ distinct weights, and obtained $\Theta(\log n)$ after rounding to powers of $2$. Their lower-bound instance is a weighted path.
Blocking, introduced through the ShortCut algorithm of Kalyanasundaram and Pruhs~\cite{kalyanasundaram1994} and reformulated by Megow et al.~\cite{megow2012}, is another central exploration algorithm.
For fixed $\delta>0$, $\Blocking$ has a competitive ratio of $\Omega(n^{1/(\delta+4)})$ on general graphs~\cite{megow2012}.
Baligács, Disser, Heinrich, and Schweitzer~\cite{baligacs2023jcss} prove an $\Omega(\log n/\log\log n)$ lower bound with adversarial ties for every choice of $\delta(n)$, and an $O(\log n)$ upper bound for $\delta=\log n$.
Their guess-and-square method for unknown $n$ motivates our phase schedules.

General deterministic lower bounds remain constant.
Birx, Disser, Hopp, and Karousatou~\cite{birx2021} prove a lower bound of~$10/3$ and, for $k\ge2$ distinct positive weights, a lower bound of~$10/3-2/(3k)$.
Baligács~\cite{baligacs2026lb} raises the lower bound to $4$ for non-negative edge weights, even on subcubic planar graphs.

Constant competitive algorithms are known for restricted graph classes. Blocking is $16$-competitive on planar graphs~\cite{kalyanasundaram1994,megow2012}.
The bounded-genus guarantee $16(1+2g)$ of Megow et al.~\cite{megow2012} was improved to $16(1+2g/3)$ by Baligács et al.~\cite{baligacs2023jcss}, who also obtain a constant competitive ratio on every class excluding a fixed minor.
\ensuremath{\DFS} has the optimal competitive ratio of $2$ on unweighted graphs~\cite{miyazaki2009} and of $1$ on trees.
On cycles, the deterministic optimum is $(1+\sqrt3)/2$~\cite{miyazaki2009}.
Baligács, Hązła, and Volk~\cite{baligacs2026rand} give a randomized algorithm with an expected competitive ratio below $1.315$ on cycles against an oblivious adversary, together with a lower bound of about $1.1155$.
Their result shows that randomization can improve exploration guarantees even in this local-information model.

Exploration with advice~\cite{brandt2020,dobrev2012}, with a predicted tour or tree~\cite{eberle2022}, and with known graph structure and uncertain edge weights~\cite{gehnen2025} provides additional information unavailable in our model.
\section{Model and preliminaries}\label{sec:model}

An \emph{instance} is a triple $(G,w,s)$, where $G=(V,E)$ is a finite, connected, undirected graph with $n=|V|\ge2$ vertices, $w\colon E\to\R_{>0}$ assigns a positive length to every edge, and $s\in V$ is the start vertex.
For $K\subseteq V$, let $G[K]$ denote the subgraph induced by $K$, with the original weights, and similarly for other graphs.
For $v\in V$, let $\delta(v)=\{e\in E:v\in e\}$ be the set of edges incident to $v$. The graph is unknown to the agent. Standing at a vertex $v$, the agent knows the identifier of $u$ and the weight of $e$ for each $e=\{v,u\}\in\delta(v)$ and recognizes vertices it has seen before by their identifiers.
Fix a total order on vertex identifiers. To break ties among eligible edges at $v$, we examine them in \emph{identifier order}, that is, in increasing order of the identifiers of their other endpoints.
A vertex is \emph{visited} once the agent has stood on it, and an edge $\{u,x\}$ with $u$ visited and $x$ unvisited is a \emph{boundary edge}.
From $v$, the agent may traverse any $e\in\delta(v)$ at cost $w(e)$. A deterministic online algorithm decides the next move from everything seen so far and does not know $n$ unless stated otherwise.

A \emph{walk} in a graph $H$ is a finite sequence $(v_0,\ldots,v_q)$ of vertices of $H$ in which consecutive vertices are joined by an edge of $H$, and its length is the total weight of its steps, counted with multiplicity. A walk is \emph{closed} if $v_q=v_0$, and a \emph{tour} of $H$ is a closed walk that visits every vertex of~$H$.
For a walk $P=(v_0,\ldots,v_q)$ and a neighbour~$x$ of $v_q$, write $P+x=(v_0,\ldots,v_q,x)$.

The task is to construct a tour of $G$ starting at $s$. We write $\ALG(G,w,s)$ for the length of the agent's tour and $\OPT(G,w,s)$ for the minimum length of such a tour. We omit the arguments when clear from context.
The algorithm is \emph{$\rho$-competitive}, in the strict sense without an additive constant, if $\ALG\le\rho\cdot\OPT$ on every instance.
For randomized online algorithms, we assume an oblivious adversary. The algorithm draws its random bits before the exploration starts and the instance is fixed in advance and does not depend on the random draws. Such an algorithm is $\rho$-competitive if $\E[\ALG]\le\rho\cdot\OPT$ on every instance.

On a tree~$T$ rooted at $r$, \emph{depth-first search} ($\DFS$) explores each child's subtree recursively, returning to the parent before taking the next child in the prescribed order. The \emph{closed $\DFS$ tour} is the vertex sequence of the resulting walk, including all backtracking steps. It starts and ends at $r$ and traverses each edge of $T$ once in each direction.

For a graph $H$ with edge weights, $\MSF(H)$ denotes the weight of a minimum spanning forest of $H$, and $\MST=\MST(G)$ the weight of a minimum spanning tree of $G$. The following lemma converts the tree-weight estimates in our analysis into bounds relative to $\OPT$. The elementary proof is given in Appendix~\ref{app:det}.

\begin{restatable}{lemma}{restateOPTbound}\label{lem:facts}
For every instance, (a) $\MST\le\frac{n-1}{n}\OPT$, and (b) $\OPT\ge2\dist(u,v)$ for all vertices $u,v$, where $\dist$ is the shortest-path distance in $(G,w)$.
\end{restatable}

We group the edges by weight, using geometric classes.

\begin{definition}\label{def:superclass}
Fix a base $b\ge2$.
\begin{enumerate}
\item[(a)] The \emph{class} of an edge $e$ is $m(e)=\lceil\log_b w(e)\rceil$, so that $b^{m(e)-1}<w(e)\le b^{m(e)}$.
\item[(b)] For an integer $m$, the graph $G_{\le m}$ has vertex set $V$ and the edges of class at most $m$. A \emph{level-$m$ cluster} is a connected component of $G_{\le m}$.
\item[(c)] For $v\in V$, $X_m(v)$ is the level-$m$ cluster containing $v$, and $N_m$ is the number of level-$m$ clusters. Finally, $m_{\max}$ is the smallest integer $m$ for which $G_{\le m}$ is connected.
\end{enumerate}
\end{definition}
The following properties are easy consequences of Kruskal's algorithm for minimum spanning trees; see Appendix~\ref{app:det} for a formal proof.

\begin{restatable}{lemma}{restateKruskal}\label{lem:kruskal}
Let $T$ be a minimum spanning tree obtained by Kruskal's algorithm and let $w_{\max}=\max_{e\in T}w(e)$. Then, for every integer $m$,
\begin{enumerate}
\item[(a)] $T\cap E(G_{\le m})$ is a minimum spanning forest of $G_{\le m}$, $T$ has exactly $N_{m-1}-N_m$ edges of class $m$, and $\MSF(G_{\le m})\le\MST$.
\item[(b)] For a level-$(m-1)$ cluster $K$, $T[K]$ is a minimum spanning tree of $G[K]$. Every minimum spanning tree of $G[K]$ uses only edges of class at most $m-1$.
\item[(c)] $m_{\max}=\lceil\log_b w_{\max}\rceil$.
\end{enumerate}
\end{restatable}
\section{The algorithm}\label{sec:algorithm}

Our algorithm $\hDFSB$ adapts the hierarchy of Megow et al.~\cite{megow2012}. At level $m$, it performs DFS between level-$(m-1)$ clusters, treating each as a vertex. Inside a fully visited cluster (one whose vertices have all been visited), the agent follows a closed DFS tour of a minimum spanning tree for the original weights. Throughout the paper, Kruskal's algorithm breaks equal-weight ties lexicographically by the pair of endpoint identifiers (smaller identifier first), and DFS visits the children of each tree vertex in identifier order.

All calls share the visited set $S$, the revealed edges $R=\bigcup_{x\in S}\delta(x)$, and the walk of the agent~$P$. For $x\in S$, let $C_m(x)$ be the set of vertices reachable from $x$ in $G[S]$ using edges of class at most $m$. The agent can compute $C_m(x)$ from $S$ and $R$. The corresponding cluster $X_m(x)$ is defined in the full graph, so $C_m(x)\subseteq X_m(x)$, with equality once the cluster is fully visited.

The main procedure (Algorithm~\ref{alg:hdfs}) starts at $s$ with the smallest class $m_0$ among the edges in $\delta(s)$. Since $X_{m_0-1}(s)=\{s\}$, its first call is $\Process{}(m_0,\{s\},s)$. It advances one level after each call. Both subroutines return the agent to their entry vertex $r$ (Lemma~\ref{lem:invariant}).

$\Process{}(m,K,r)$ (Algorithm~\ref{alg:process}) starts in a fully visited cluster $K=X_{m-1}(r)$, with every other level-$(m-1)$ cluster either fully visited or unvisited. If a class-$m$ boundary edge leaves $K$, the agent starts its tree tour. On first reaching a tour vertex, it examines its class-$m$ boundary edges in identifier order. Crossing such an edge to $x$ is a \emph{forward traversal}. This is the only step after initialization that adds a vertex to $S$ and its incident edges to $R$. The agent calls $\Explore{}(m-1,x)$ to complete the new cluster $K'$, then $\Process{}(m,K',x)$ to search from it. It crosses back along the same edge (a \emph{backward traversal}) and resumes the interrupted tour.

$\Explore{}(m,r)$ (Algorithm~\ref{alg:explore}) starts at the only visited vertex $r$ of $X_m(r)$. Unless $X_m(r)=\{r\}$, it first completes $K=X_{m-1}(r)$ recursively. The call $\Process{}(m,K,r)$ then visits the rest of $X_m(r)$.

\begin{algorithm}[tb]
\small
\caption{$\hDFSB$}\label{alg:hdfs}
\KwIn{Start vertex $s$ and base $b\ge2$.}
\KwOut{A closed walk $P$ from $s$ visiting every vertex.}
$S\gets\{s\}$, $R\gets\delta(s)$, $P\gets(s)$\;
$m\gets\min\{m(e):e\in\delta(s)\}$\;
\While{there is a boundary edge}{
  $\Process{}(m,C_{m-1}(s),s)$\;
  $m\gets m+1$\;
}
\Return{$P$}\;
\end{algorithm}
\begin{algorithm}[tb]
\small
\caption{\Process{}($m,K,r$)}\label{alg:process}
\KwIn{Level $m$, fully visited level-$(m-1)$ cluster $K$, and current vertex $r\in K$.}
\KwOut{Updates $S$, $R$, and~$P$; $P$ ends in~$r$ and both endpoints of every class-$m$ edge incident to $K$ are visited.}
\lIf{no class-$m$ boundary edge leaves $K$}{\Return}
$T\gets$ Kruskal tree of $G[K]$ for the original weights\;
$(v_0,\ldots,v_q)\gets$ closed \ensuremath{\DFS} tour of $T$ starting at $r$\;
\For{$j=0,\ldots,q$}{
  \lIf{$j>0$}{$P\gets P+v_j$\tcp*[f]{travel}} 
  \If{$v_j\notin\{v_i:0\le i<j\}$}{
    \ForEach{class-$m$ edge $\{v_j,x\}\in\delta(v_j)$ in identifier order}{
      \If{$x\notin S$}{
        $P\gets P+x$, $S\gets S\cup\{x\}$, $R\gets R\cup\delta(x)$\tcp*{forward traversal}
        $K'\gets\Explore{}(m-1,x)$\;
        $\Process{}(m,K',x)$\;
        $P\gets P+v_j$\tcp*{backward traversal}
      }
    }
  }
}
\end{algorithm}

\begin{algorithm}[tb]
\small
\caption{\Explore{}($m,r$)}\label{alg:explore}
\KwIn{Level $m$ and current vertex $r$, the only visited vertex of $X_m(r)$.}
\KwOut{Returns $X_m(r)$. Updates $S$, $R$, and $P$. On return, $X_m(r)\subseteq S$, $R$ contains all edges incident to vertices of $X_m(r)$, and $P$ ends at $r$.}
\lIf{$m(e)>m$ for all $e\in\delta(r)$}{\Return{$\{r\}$}}
$K\gets\Explore{}(m-1,r)$\;
$\Process{}(m,K,r)$\;
\Return{$C_m(r)$}\tcp*{now equal to $X_m(r)$}
\end{algorithm}

Every boundary edge has a visited endpoint, so the agent can test for one in $R$. Since $G$ is connected, the main loop finds no boundary edge exactly when $S=V$, even if $n$ is unknown. A level-$m$ call on $K$ needs to test only the edges incident to $K$, all of which are known. If no class-$m$ boundary edge leaves $K$, the call returns without moving. Once a tree tour starts, it runs to completion even if the last vertex is visited along the way. The tightness results use this convention.\label{rem:online}
\section{Analysis of the deterministic algorithm}\label{sec:analysis}

Fix $b\ge2$ and a minimum spanning tree $T$ computed by Kruskal's algorithm with the same tie order as \Process{}. For every level-$m$ cluster $K$, the local tree computed on $G[K]$ is exactly $T[K]$. Indeed, by the end of class $m$, Kruskal's algorithm has connected $K$ before considering any edge leaving it. Up to that point the local and global runs examine the same internal edges in the same order and make identical decisions inside $K$.
Let $w_{\max}=\max_{e\in T}w(e)$.

At any point in the run, $S$ is the set of vertices the agent has occupied so far. We call $S$ \emph{closed at level $m$} if every level-$m$ cluster containing a vertex of~$S$ lies entirely in~$S$.
To analyse $\Process{}(m,K,r)$, contract each level-$(m-1)$ cluster of $G_{\le m}$ to a vertex and discard self-loops (Megow et al.~\cite{megow2012} describe $\hDFS$ as a depth-first search in such a contracted graph). In this \emph{quotient graph}, the edges are exactly the class-$m$ edges between distinct clusters, and each connected component represents a level-$m$ cluster.

The proof idea for Lemma~\ref{lem:invariant} is to view the level-$m$ calls to \Process{} as a depth-first search from $K$ in this graph. Mark each fully visited cluster. When \Process{} crosses a boundary edge to $x$, closure ensures that the cluster containing~$x$ was entirely unvisited. By induction, $\Explore{}(m-1,x)$ completes that cluster, so it can be marked before the next level-$m$ call searches from it. This restores closure for the next forward traversal. Appendix~\ref{app:det} gives the formal induction.

\begin{restatable}[Invariants]{lemma}{restateInvariant}\label{lem:invariant}
The following statements hold for Algorithm~\ref{alg:hdfs}.
\begin{enumerate}
\item[(a)] Every call $\Process{}(m,K,r)$ starts at $r\in K$ with $S$ closed at level $m-1$. The call returns to $r$, stays inside $X_m(r)$, and leaves every level-$(m-1)$ cluster adjacent to $K$ by a class-$m$ edge fully visited. When this call does a forward traversal, $S$ is closed at level $m-1$, so the traversal enters a previously unvisited level-$(m-1)$ cluster.
\item[(b)] A call $\Explore{}(m,r)$ starts with $r$ as the only visited vertex of $X_m(r)$. At that moment, every other level-$m$ cluster is either entirely unvisited or fully visited. It visits all of $X_m(r)$, stays inside this cluster, and returns to $r$.
\item[(c)] At the beginning of top-level iteration $m$, $S=X_{m-1}(s)$ and the agent is at $s$. At its end $S=X_m(s)$ and the agent is again at $s$.
\end{enumerate}
\end{restatable}

By Lemma~\ref{lem:invariant}(a), each cluster is toured at most once per level.

\begin{corollary}\label{cor:once}
For every integer $m$ and every level-$(m-1)$ cluster $K$, at most one call $\Process{}(m,K,r)$ starts a tree tour of $K$. If such a tour starts, then $m\le m_{\max}$.
\end{corollary}
\begin{proof}
Before $\Process{}(m,K,r)$ returns, each recursive call at level $m$ starts in a cluster that was unvisited just before the forward traversal to it (Lemma~\ref{lem:invariant}(a)). Since $K$ was already fully visited when the original call began, none of these recursive calls starts in $K$. After the call returns, all its quotient neighbours are visited by Lemma~\ref{lem:invariant}(a). A later call on $K$ therefore returns at its initial boundary-edge test.
When the tour starts, a class-$m$ boundary edge leaves $K$, so $G_{\le m-1}$ is disconnected and $m\le m_{\max}$.\qed
\end{proof}

\subsection{Accounting of costs}

We bound the cost of any prefix of the walk because a phase of the unknown-size algorithm can stop in the middle of a recursive call (Section~\ref{sec:unknown-size}). Lemma~\ref{lem:cost}(a) bounds the cost of forward and backward traversals between clusters. Parts (b) and (c) bound the cost of travel within clusters.

\begin{lemma}\label{lem:cost}
Fix a prefix of the run. For $e\in T$, let $r_e$ count the levels at which \Process{} has started a tree tour of a cluster containing both endpoints of $e$. Then
\begin{enumerate}
\item[(a)] the total cost of level-$m$ forward and backward traversals made during the prefix is at most $2(N_{m-1}-N_m)b^m$,
\item[(b)] total travel costs at most $2\sum_{e\in T}w(e)r_e$,
\item[(c)] $r_e\le m_{\max}-m(e)\le\lceil\log_b(w_{\max}/w(e))\rceil$. Moreover, $r_e>0$ implies that both endpoints of $e$ are visited.
\end{enumerate}
\end{lemma}
\begin{proof}
We first prove (a). Fix a level $m$. Each forward traversal at this level enters a previously unvisited level-$(m-1)$ cluster (Lemma~\ref{lem:invariant}(a)). If a level-$m$ cluster contains $c$ such clusters, at most $c-1$ can be entered this way, since the first traversal starts in one of them. Summing over the $N_m$ level-$m$ clusters gives at most $N_{m-1}-N_m$ forward traversals. Every backward traversal in the prefix matches a distinct earlier forward traversal in the same prefix. Since each crossing costs at most $b^m$, their total cost is at most $2(N_{m-1}-N_m)b^m$, which proves (a).

For (b), charge the full tour of $T[K]$ when \Process{} starts it, even if the prefix ends before the tour does. Each tree edge~$e$ contributes~$w(e)$ in each direction, and Corollary~\ref{cor:once} allows at most one such charge per level. This proves (b).

For (c), note that a tour at level $m$ uses only edges of class less than $m$. Moreover, the main loop terminates by level $m_{\max}$, since at the end of that iteration all vertices have been visited (Lemma~\ref{lem:invariant}(c)). Thus $e$ can be charged only at levels $m(e)+1,\ldots,m_{\max}$. Substituting $m_{\max}=\lceil\log_b w_{\max}\rceil$ and $m(e)=\lceil\log_b w(e)\rceil$ and using $\lceil x\rceil-\lceil y\rceil\le\lceil x-y\rceil$ for real $x,y$ gives the two inequalities in (c). Both endpoints of a charged edge have already been visited because $K$ is fully visited when the tour starts.\qed
\end{proof}

We can now prove the main result of the paper.

\begin{theorem}\label{thm:main}
At any moment during a run of $\hDFSB$, let $n'=|S|$ be the number of distinct vertices visited so far, including the start vertex $s$. The cost incurred by that moment is at most $\phi\,\MST\le\frac{n-1}{n}\phi\,\OPT$, where $\phi=2b+2+2\log_b\min\{n',n-1\}$.
When $n$ is known, choosing $b=\max\{2,\ln n/(\ln\ln n)^2\}$ gives a competitive ratio of ${(2+o(1))}\ln n/\ln\ln n$.
\end{theorem}
\begin{proof}
Summing Lemma~\ref{lem:cost}(a) over all classes and applying Lemma~\ref{lem:kruskal}(a) bounds the cost of traversals by
\[
2\sum_m(N_{m-1}-N_m)b^m
=2\sum_{e\in T}b^{m(e)}\le2b\,\MST.
\]
Let $T'=\{e\in T:r_e>0\}$ and choose $f\in T$ of weight $w_{\max}$.
By Lemma~\ref{lem:cost}(c), $T'$ is a forest on the $n'$ visited vertices and $f\notin T'$. Hence $T^+=T'\cup\{f\}$ has at most $\min\{n',n-1\}$ edges, maximum weight $w_{\max}$, and total weight at most $\MST$.
Put $M=\sum_{e\in T^+}w(e)$. Applying Jensen's inequality to the logarithm with weights $w(e)/M$ gives
\[
\frac1M\sum_{e\in T^+}w(e)\ln\biggl(\frac{w_{\max}}{w(e)}\biggr)
\le \ln\Biggl(\sum_{e\in T^+}\frac{w(e)}M\frac{w_{\max}}{w(e)}\Biggr)
=\ln \biggl(\frac{|T^+|w_{\max}}M \biggr)\le\ln |T^+|.
\]
The total travel cost is therefore at most
\begin{align*}
2\sum_{e\in T'}w(e)\biggl(1+\log_b\frac{w_{\max}}{w(e)}\biggr)
&=2w(T')+\frac2{\ln b}\sum_{e\in T^+}w(e)\ln \biggl(\frac{w_{\max}}{w(e)}\biggr)\\
&\le2(\MST-w_{\max})+2\MST\log_b\min\{n',n-1\}.
\end{align*}
Adding the two costs gives the bound $\phi\,\MST$, and Lemma~\ref{lem:facts}(a) gives the bound in terms of $\OPT$.
At termination $n'=n$, so the logarithmic term becomes $2\log_b(n-1)$.

For sufficiently large $n$, the chosen base is $b=\ln n/(\ln\ln n)^2$. To compare the terms in the bound, divide each by $\ln n/\ln\ln n$. The contribution from $2b+2$ vanishes, since
\[
\frac{2b+2}{\ln n/\ln\ln n}
=\frac{2}{\ln\ln n}+\frac{2\ln\ln n}{\ln n}\longrightarrow 0.
\]
For the logarithmic term, $\ln b=\ln\ln n-2\ln\ln\ln n$, so
\[
\frac{2\log_b(n-1)}{\ln n/\ln\ln n}
=2\,\frac{\ln(n-1)}{\ln n}\,
\frac{1}{1-2\ln\ln\ln n/\ln\ln n}\longrightarrow 2.
\]
Indeed, $\ln(n-1)/\ln n\to1$ and $\ln\ln\ln n/\ln\ln n\to0$. Since the prefactor $(n-1)/n$ also tends to $1$, the competitive ratio is at most $(2+o(1))\ln n/\ln\ln n$.\qed
\end{proof}

\subsection{Known range of weights}

Suppose that every edge weight lies in $[1,2^k]$. Rounding each weight up to a power of two leaves at most $1+\lceil k\rceil$ distinct weights, does not decrease the cost of any walk, and at most doubles $\OPT$. Hence the $2d$ bound of~\cite[Theorem~4]{megow2012} for $d$ distinct weights gives power-of-two $\hDFS$ a competitive ratio of at most $4(1+\lceil k\rceil)$. The path with weights $1,2,\ldots,2^{d-1}$ in the proof of~\cite[Theorem~5]{megow2012}, with $d=\lfloor k\rfloor+1$, shows that this linear dependence on $k$ is tight for that algorithm.
For $\hDFSB$, increasing $b$ reduces the number of levels at which travel inside a cluster can incur cost. When $k$ is known, we can use it to choose the base without knowing $n$.

\begin{theorem}\label{thm:range}
If all edge weights lie in $[1,2^k]$ for some $k\ge0$, then for every $b\ge2$ the competitive ratio of $\hDFSB$ is at most $\frac{n-1}{n}(2b+2\lceil k/\log_2 b\rceil)$.
Choosing $b=k/\log_2 k$ for $k\ge4$ and $b=2$ for $0\le k<4$ gives a competitive ratio of $O(k/\log k)$ as $k\to\infty$ and a constant bound for bounded $k$.
\end{theorem}
\begin{proof}
Let~$\bar m=\lceil\log_b(2^k)\rceil=\lceil k/\log_2 b\rceil$. Then, every edge has a class in $\{0,\dots,\bar m\}$, so $G_{\le\bar m}=G$ and $m_{\max}\le\bar m$. By Lemma~\ref{lem:cost}(c), the number of levels charging a tree edge $e$ satisfies $r_e\le m_{\max}-m(e)\le\bar m$. Lemma~\ref{lem:cost}(b) therefore bounds total travel by $2\sum_{e\in T}\bar m\,w(e)=2\bar m\MST$.

For forward and backward traversals, Lemmas~\ref{lem:cost}(a) and~\ref{lem:kruskal}(a) give a total cost of at most $2\sum_{e\in T}b^{m(e)}$. Since $b^{m(e)-1}<w(e)$, this is at most $2b\MST$. Adding travel and applying Lemma~\ref{lem:facts}(a) gives
\[
\ALG\le (2b+2\bar m)\MST\le\frac{n-1}{n}(2b+2\bar m)\OPT.
\]

For $0\le k<4$, choosing $b=2$ gives $\bar m=\lceil k\rceil\le4$ and hence a competitive ratio of at most $12$. For $k\ge4$, choose $b=k/\log_2 k$. This satisfies the algorithm's requirement $b\ge2$ because $k/\log_2 k$ equals $2$ at $k=4$ and increases thereafter. With this choice,
$\log_2 b=\log_2 k-\log_2\log_2 k=(1-o(1))\log_2 k$.
Thus $\bar m\le k/\log_2 b+1=O(k/\log k)$ as $k\to\infty$. Both terms in $2b+2\bar m$ have the claimed order.\qed
\end{proof}
\section{Randomizing the weight classes}\label{sec:random}

A standard way to randomize geometric thresholds $b^j$ is to replace them by~$b^{j+u}$, using a single offset $u$ drawn uniformly from $[0,1)$. For any fixed weight, the distance on the logarithmic scale to the next threshold is then uniform on~$[0,1)$; see, e.g., \cite{chrobak2008,griesbach2026,kao1996} for further applications of this technique.

Here the thresholds define the weight classes. For $u\in\R$, let $\hDFSBu$ be Algorithm~\ref{alg:hdfs} with $m_u(e)=\lceil\log_bw(e)-u\rceil$. We fix the instance and average over the uniform offset~$u$ drawn before exploration. For known $n$, this sharpens the finite bound in Theorem~\ref{thm:main}, while the leading term remains $2\ln n/\ln\ln n$.

Our main result of this section is the following; it is proven in Appendix~\ref{app:offset} and follows the standard framework for analysing such random offsets.

\begin{restatable}{theorem}{restateOffset}\label{thm:offset}
For every $b\ge2$ and every $n$-vertex instance with $n\ge2$,
\[
\E_u[\ALG(\hDFSBu)]\le
\frac{n-1}{n}\frac{2(b-1+\ln(n-1))}{\ln b}\OPT.
\]
For known $n$, the choice $b=\max\{2,\ln n/(\ln\ln n)^2\}$ gives an expected competitive ratio of ${(2+o(1))}\ln n/\ln\ln n$.
\end{restatable}

The following theorem shows that the analysis in Theorems~\ref{thm:main} and~\ref{thm:offset} is tight. It summarizes Theorems~\ref{thm:tight} and~\ref{thm:random-tight}, which are proved in Appendix~\ref{sec:tight}.

\begin{theorem}\label{thm:lower}
\begin{enumerate}
\item[(a)] For every $n\ge16$ and every base $b\ge2$, which may depend on $n$, the competitive ratio of $\hDFSB$ on $n$-vertex instances is at least $\log_2n/\log_2\log_2n-5$.
\item[(b)] Let $b=\max\{2,\ln n/(\ln\ln n)^2\}$. For every sufficiently large $n$, there is an $n$-vertex instance on which $\ALG(\hDFSBu)\ge(2-o(1))(\ln n/\ln\ln n)\OPT$ for every $u\in[0,1)$. In particular, the bound holds for $\hDFSB$ and for $\E_u[\ALG(\hDFSBu)]$ with $u$ uniform on $[0,1)$.
\end{enumerate}
\end{theorem}

\section{Unknown number of vertices}\label{sec:unknown-size}

Baligács~\cite[Proposition~2]{baligacs2026lb} shows that knowing $n$ does not change the best achievable constant competitive ratio. Its proof changes $n$, so the argument does not apply to ratios that grow with $n$.
Our (deterministic and randomized) algorithms choose the base~$b$ depending on the graph size (Theorems~\ref{thm:main} and \ref{thm:offset}). When $n$ is unknown, this is not possible. Instead, we run the exploration in phases (cf.~Algorithm~\ref{alg:phases}). Each phase~$i$ guesses a number of vertices $\hat n_i$ and chooses its base as if that guess were the graph size.

In phase $i=0,1,\ldots$, the algorithm computes its guess for $\ln n$ as $\smash{y_i=t_i^{t_i/2}}$. The deterministic schedule uses $t_i=i+2$. The randomized schedule draws a single offset $\nu$ uniformly from~$[0,1)$ before exploration and uses $t_i=i+2+\nu$ in every phase. Exponentiating gives the vertex budget $\hat n_i=\lfloor e^{y_i}\rfloor$, and substituting $y_i$ for $\ln n$ in the known-size base gives $b_i$. The first phase with $y_i\ge\ln n$ has enough budget to finish. 
The choice $\smash{y_i=t_i^{t_i/2}}$ balances the cost of restarting against the cost of guessing too high. Consecutive guesses for $\ln n$ differ by a factor of order $\sqrt{t_i}$. This factor grows, so the last failed guess dominates the sum of all failed guesses. With a uniform shift, even that last guess is unlikely to be close to $\ln n$. Only a shrinking fraction of the possible shifts place it within a fixed factor of $\ln n$. Yet the factor grows slowly enough that the first successful phase still uses a base $b_i=o(\ln n/\ln\ln n)$, as shown in Appendix~\ref{app:phases}. Thus the cost of crossings in that phase remains a lower-order term in the competitive bound. 

If a phase exceeds its guess at vertex $v$, the agent stops that run immediately and starts the next phase at $v$. The new run initializes its visited set to $S=\{v\}$ and its revealed edges to $R=\delta(v)$. It treats every other vertex as unvisited, including vertices reached in earlier phases, and uses only the edges revealed in the current phase. Thus $v$ counts as the first vertex of the new phase. Each other vertex counts towards the new guess on its first visit in this phase, however often the agent visited it before.

\begin{algorithm}[tb]
\small
\caption{Algorithms for unknown $n$}\label{alg:phases}
\KwIn{Start vertex $s$.}
\KwOut{A closed exploration walk from $s$.}
\textit{randomized algorithm:} draw $\nu$ uniformly from $[0,1)$\;
\textit{deterministic algorithm:} set $\nu\gets0$\;
\For{$i=0,1,\ldots$}{
  $t_i\gets i+2+\nu$, $y_i\gets t_i^{t_i/2}$, $\hat n_i\gets\lfloor e^{y_i}\rfloor$, $b_i\gets\max\{2,y_i/(\ln y_i)^2\}$\;
  $v\gets$ current vertex\;
  run $\hDFSs_{b_i}$ from $v$ ($S=\{v\}$, $R=\delta(v)$), and abort it as soon as $|S|=\hat n_i+1$\;
  \lIf{the run was aborted}{\Continue}
  return to $s$ along a shortest path and \textbf{stop}\;
}
\end{algorithm}

The walks of all phases concatenate into a single walk, and every step counts towards $\ALG$. The successful phase visits every vertex during that phase and ends at its own start vertex. The agent then knows the whole graph and can take a shortest path back to the original start $s$. We prove the following in Appendix~\ref{app:phases}.

\begin{restatable}{theorem}{restatePhases}\label{thm:phases}
Without knowing $n$, Algorithm~\ref{alg:phases} satisfies, as $n\to\infty$,
\begin{enumerate}
\item[(a)] $\E_\nu[\ALG]\le(2+o(1))(\ln n/\ln\ln n)\OPT$ for the randomized algorithm,
\item[(b)] $\ALG\le(4+o(1))(\ln n/\ln\ln n)\OPT$ for the deterministic algorithm.
\end{enumerate}
\end{restatable}
\section{Open problems}\label{sec:open}

Does online exploration of general graphs admit a constant-competitive deterministic algorithm?
Our upper bound narrows the gap, while the lower bounds in this paper apply only to the specified $\hDFSB$ family.
Whether a competitive ratio of $o(\log n/\log\log n)$ is achievable, deterministically or against an oblivious adversary, remains open.
It would also be useful to obtain the bounded-weight-range guarantee without knowing the range.%
\label{page:end-of-main-text}

\section*{Acknowledgements}
Large language models were used in preparing this work. Claude Code (Anthropic) assisted with literature search, code and experiments, proof development and checking, and drafting. ChatGPT and Codex (OpenAI) assisted with research notes, algorithm and proof development, manuscript review, proof simplification, and editing. The author is responsible for the correctness and originality of the results and for the final text.

\bibliographystyle{splncs04}
\bibliography{refs}

\appendix
\clearpage
\section{Proofs of the structural lemmas}\label{app:det}

We prove the structural facts used in Section~\ref{sec:analysis}.
Throughout, $T$ is a minimum spanning tree obtained by Kruskal's algorithm with the tie order used by \Process{}, and $w_{\max}=\max_{e\in T}w(e)$.

\restateOPTbound*
\begin{proof}
List the $n$ vertices in order of first visit on an optimal tour. Join consecutive vertices in this cyclic order by shortest paths, including the last vertex back to the first. Their total length is at most $\OPT$. Removing a longest path from this list leaves $n-1$ paths of total length at most $(n-1)\OPT/n$ whose union connects all vertices. This proves (a). For (b), the two portions of the optimal tour between $u$ and $v$ each have length at least $\dist(u,v)$.\qed
\end{proof}

\subsection{Kruskal's algorithm and the clusters}

\restateKruskal*
\begin{proof}
Kruskal processes all edges of $G_{\le m}$ before any heavier edge, so its accepted edges at that point form a minimum spanning forest of $G_{\le m}$, with $n-N_m$ edges. Write $F_j=T\cap E(G_{\le j})$ for this forest at threshold $j$, so $|F_j|=n-N_j$. Since $F_{m-1}\subseteq F_m$, the class-$m$ edges of $T$ are precisely those in $F_m\setminus F_{m-1}$. Their number is
\[
|F_m|-|F_{m-1}|=(n-N_m)-(n-N_{m-1})=N_{m-1}-N_m.
\]
Moreover, $F_m\subseteq T$ gives $\MSF(G_{\le m})\le\MST$, completing (a). The forest first becomes connected when its heaviest edge is accepted, which proves (c).

For (b), the restriction of the forest at threshold $m-1$ spans $K$ and is minimum there. No later edge of $T$ can have both endpoints in $K$, since it would create a cycle. Finally, a spanning tree of $G[K]$ containing an edge heavier than $b^{m-1}$ can be improved by removing that edge and reconnecting its two components along a path in $G_{\le m-1}[K]$.\qed
\end{proof}

\subsection{The quotient \ensuremath{\DFS} invariant}

\restateInvariant*
\begin{proof}
\emph{Induction statement.} Let $I(m)$ be the conjunction of the following two implications. We prove $I(m)$ by induction on $m$.
\begin{itemize}
\item For $\Process{}(m,K,r)$, assume that the agent is at $r$ and that $K=X_{m-1}(r)$ is the level-$(m-1)$ cluster containing $r$. Assume also that $K\subseteq S$ (all vertices of $K$ have been visited) and that $S$ is closed at level $m-1$. Under these entry conditions, the call performs a depth-first search from $K$ in the quotient graph, treating the other fully visited clusters as already marked. It satisfies the return and traversal guarantees in (a) and leaves $S$ closed at level $m-1$.
\item For $\Explore{}(m,r)$, assume that the agent is at $r$, that $S\cap X_m(r)=\{r\}$, and that every other level-$m$ cluster is fully visited or unvisited. Under these entry conditions, the call visits all of $X_m(r)$, stays within it, and returns its vertex set with the agent back at $r$.
\end{itemize}
If $m$ is below the smallest edge class, every cluster at that level is a singleton. Both calls return immediately, proving the base case.

\emph{Induction step for \Process{}.} Assume $I(m-1)$. We first prove the implication for \Process{} in $I(m)$, using the implication for \Explore{} in $I(m-1)$. For the recursive calls at the same level, we use an inner induction on the number of unvisited level-$(m-1)$ clusters in $X_m(r)$ when the call starts. A nested call $\Process{}(m,K',x)$ starts only after $\Explore{}(m-1,x)$ has filled $K'$, so this number is smaller for the nested call, and the inner induction hypothesis applies to it. When $\Process{}(m,K,r)$ begins, $S$ is closed at level $m-1$, so each level-$(m-1)$ cluster is either fully visited or entirely unvisited. In the quotient graph, mark the vertex of each fully visited cluster, including $K$. Consider a forward traversal to $x$ while this marking represents $S$. Since $x$ is unvisited, its cluster $K'=X_{m-1}(x)$ is unmarked and entirely unvisited. Immediately after the traversal, $x$ is its only visited vertex and every other level-$(m-1)$ cluster is still fully visited or unvisited. These are precisely the hypotheses for $\Explore{}(m-1,x)$. By induction, that call fills $K'$, stays inside it, and returns to $x$. We can now mark $K'$: closure has been restored before $\Process{}(m,K',x)$ starts.

The same-level calls therefore follow ordinary DFS on the quotient: each recursive call starts at a newly marked vertex, so the recursion terminates. The tour of $K$ examines every incident class-$m$ edge unless all neighbours were already marked at entry. Hence all neighbours are marked on return. Each recursive excursion returns along its entering edge, and the closed tour of $K$ ends at $r$. All moves stay within the quotient component of $K$, that is, within $X_m(r)$. This proves the first implication of $I(m)$.

\emph{Induction step for \Explore{}.} The singleton case is immediate. Otherwise let $K=X_{m-1}(r)$. Since $K\subseteq X_m(r)$, only $r$ has been visited in $K$. The other level-$(m-1)$ clusters inside $X_m(r)$ are unvisited, and those outside it lie in fully visited or unvisited level-$m$ clusters. Hence the induction hypothesis applies to $\Explore{}(m-1,r)$: it fills $K$ and returns to $r$. Now $S$ is closed at level $m-1$, and $K$ is the only marked vertex in its quotient component. By the first claim at level $m$, $\Process{}(m,K,r)$ searches that entire component and returns to $r$. Its clusters together form $X_m(r)$, so $C_m(r)=X_m(r)$ at return. This proves the second claim.

\emph{Top-level iterations.} Initially $S=\{s\}=X_{m_0-1}(s)$. Suppose iteration $m$ starts at $s$ with $S=X_{m-1}(s)$. This is a single fully visited quotient vertex, so the first claim applies and the search visits exactly its component, $X_m(s)$, returning to $s$. This proves (c) and supplies the entry state for the next iteration. The induction steps above verified the entry conditions for each recursive call, so (a) and (b) hold for every call made by the algorithm.\qed
\end{proof}
\section{Proof of the random-offset bound}\label{app:offset}

\restateOffset*
\begin{proof}
Fix an instance, a base $b\ge2$, a Kruskal minimum spanning tree $T$, and an edge $f\in T$ of maximum weight $w_{\max}$.
For an offset $u$ write
\[
m_u(e)=\lceil\log_bw(e)-u\rceil,
\qquad q_u(e)=b^{m_u(e)+u}.
\]
Multiplying all weights by $b^{-u}$ gives precisely the classes $m_u$. This scaling preserves the order of edges, including the fixed tie order, so Kruskal's algorithm gives the same tree $T$ for every $u$. The run of $\hDFSBu$ on $(G,w)$ makes the same moves as the run of $\hDFSB$ on $(G,b^{-u}w)$, since every decision depends only on the classes, the identifiers, and the order of the weights. We can therefore apply the deterministic cost bounds to the scaled instance and multiply costs by $b^u$ to recover the original units.

Fix $u$. By Lemmas~\ref{lem:cost}(a) and~\ref{lem:kruskal}(a), the cost of forward and backward traversals in each class is at most twice its upper weight threshold times the number of tree edges in that class. Each $e\in T$ thus contributes $2q_u(e)$ to this bound. For travel within clusters, the highest relevant level is $m_u(f)$ by Lemma~\ref{lem:kruskal}(c). A tree edge $e$ can be charged $2w(e)$ at most once at each level from $m_u(e)+1$ through $m_u(f)$ (Lemma~\ref{lem:cost}(b,c)). Adding the two bounds gives
\begin{equation}\label{eq:offset-fixed-u}
\ALG(\hDFSBu)\le
2\sum_{e\in T}q_u(e)
+2\sum_{e\in T}w(e)\bigl(m_u(f)-m_u(e)\bigr).
\end{equation}
For $u$ uniform on $[0,1)$ and every real $x$, we have
\begin{equation}\label{eq:offset-identities}
\E \bigl[ \lceil x-u\rceil \bigr]=x,
\qquad
\E \bigl[b^{\lceil x-u\rceil+u}\bigr] =b^x\frac{b-1}{\ln b}.
\end{equation}
Indeed, if $x=k+t$ with $k\in\mathbb Z$ and $t\in[0,1)$, then $\lceil x-u\rceil$ equals $k+1$ with probability $t$ and $k$ otherwise. Also $\lceil x-u\rceil-(x-u)$ is the fractional part of $u-x$, which is uniform on $[0,1)$, and $\int_0^1b^z\,\mathrm{d}z=(b-1)/\ln b$.

Applying~\eqref{eq:offset-identities} with $x=\log_b w(e)$ gives $\E \bigl[q_u(e) \bigr]=w(e)(b-1)/\ln b$. By linearity of expectation, the expected class difference $\E\bigl[m_u(f)-m_u(e)\bigr]$ equals $\log_b(w_{\max}/w(e))$, even though both classes use the same offset. Taking expectations in~\eqref{eq:offset-fixed-u} yields
\[
\E\bigl[\ALG\bigr]\le \frac{2}{\ln b}\left((b-1)\MST+
\sum_{e\in T}w(e)\ln \biggl(\frac{w_{\max}}{w(e)}\biggr)\right).
\]
These expectations exist because the run is constant between the finitely many offsets at which an edge changes class.

To bound the sum, apply Jensen's inequality for $\ln$ with weights $p_e=w(e)/\MST$. Since $T$ has $n-1$ edges and $w_{\max}\le\MST$,
\[
\sum_{e\in T}p_e\ln \biggl(\frac{w_{\max}}{w(e)}\biggr)
\le \ln\!\left(\sum_{e\in T}p_e\frac{w_{\max}}{w(e)}\right)
=\ln \biggl(\frac{(n-1)w_{\max}}{\MST} \biggr) \le\ln(n-1).
\]
Substituting this estimate and using $\MST\le (n-1)\OPT/n$ from Lemma~\ref{lem:facts}(a) gives the stated finite bound.

For sufficiently large $n$, the chosen base is $b=\ln n/(\ln\ln n)^2$. Since $\ln(n-1)=\ln n+O(1/n)$, we have
\[
b-1+\ln(n-1)=\ln n\left(1+O\!\left(\frac{1}{(\ln\ln n)^2}\right)\right).
\]
Also $\ln b=\ln\ln n-2\ln\ln\ln n$. Factoring out $\ln\ln n$ and using $1/(1-x)=1+O(x)$ as $x\to0$ gives
\[
\frac{1}{\ln b}=\frac{1}{\ln\ln n}
\left(1+O\!\left(\frac{\ln\ln\ln n}{\ln\ln n}\right)\right).
\]
Substituting into the finite bound and using $(n-1)/n\le1$ gives an expected competitive ratio of at most
$\bigl(2+O(\ln\ln\ln n/\ln\ln n)\bigr)\ln n/\ln\ln n$.
The error term tends to zero, as claimed.\qed
\end{proof}
\needspace{10\baselineskip}
\section{Proofs for unknown $n$}\label{app:phases}

All asymptotic estimates below refer to $n\to\infty$ and are uniform in the schedule shift $\nu\in[0,1)$.

\restatePhases*
\begin{proof}
Let $\Lambda=2\ln n/\ln\ln n$ and $h(z)=z\ln z/2$. Fix the shift $\nu$, so that phase $i$ has parameters $t_i=i+2+\nu$ and $y_i=e^{h(t_i)}$. For sufficiently large $n$, there is a unique $\hat{t}\ge2$ with $h(\hat{t})=\ln\ln n$. Thus $e^{h(\hat{t})}=\ln n$, and $\hat{t}=O(\ln\ln n)$.

Each phase uses a fresh visited set and has the same $\OPT$, since an optimal tour can start at any vertex. The phase aborts if $n>\hat n_i$ and otherwise completes a closed tour from its own start vertex. Since
\[
\hat n_i=\lfloor e^{y_i}\rfloor\ge n
\quad\Longleftrightarrow\quad y_i\ge\ln n
\quad\Longleftrightarrow\quad t_i\ge \hat{t},
\]
the first successful phase has index $i^*=\min\{i\ge0:t_i\ge\hat{t}\}$ and parameter $t^*=t_{i^*}\in[\hat{t},\hat{t}+1)$. Since $\hat{t}$ is determined by $n$ and $t_i=i+2+\nu$, the index $i^*$ depends only on $n$ and $\nu$. Write $y^*=y_{i^*}=e^{h(t^*)}$ for its guess for $\ln n$ and $b^*=b_{i^*}$ for its base.

We first count the total cost of the successful phase with index $i^*$.
Since $t^* \in [\hat{t},\hat{t}+1)$ and $h'(z)=(\ln z+1)/2$ is increasing, and $h(\hat{t}) = \ln \ln n$, the mean value theorem gives $0\le h(t^*)-\ln\ln n\le h'(\hat{t}+1)$. Exponentiating the resulting inequality $\ln \ln n \leq h(t^*) \leq h'(\hat{t}+1) + \ln \ln n$ yields
\[
\ln n\le y^*\le (\ln n)\sqrt{e(\hat{t}+1)}
=O\bigl((\ln n)\sqrt{\ln\ln n}\bigr).
\]
For sufficiently large $n$, the base is $b^*=y^*/(\ln y^*)^2$. The lower and upper bounds on $y^*$ give
\[
b^*=O\!\left(\frac{\ln n}{(\ln\ln n)^{3/2}}\right)=o(\Lambda),
\qquad \ln b^*=(1+o(1))\ln\ln n.
\]
The successful phase visits all $n$ vertices. By Theorem~\ref{thm:main}, its cost divided by $\OPT$ is at most
\begin{equation}\label{eq:ph-success}
2b^*+2+\frac{2\ln n}{\ln b^*}=(1+o(1))\Lambda.
\end{equation}

We next count the costs of the failed phases.
For a failed phase with $y_i=y$, the base is $b(y)=\max\{2,y/(\ln y)^2\}$ and the phase stops at its $(\lfloor e^y\rfloor+1)$st distinct vertex. By Theorem~\ref{thm:main}, its cost divided by $\OPT$ is at most
\[
Q(y)=2b(y)+2+\frac{2\ln(\lfloor e^y\rfloor+1)}{\ln b(y)}.
\]
For large $y$, $b(y)=y/(\ln y)^2$, so $2b(y)+2=o(y/\ln y)$ and $\ln b(y)=\ln y-2\ln\ln y=(1+o(1))\ln y$. Also $e^y\le\lfloor e^y\rfloor+1\le e^y+1$, giving $\ln(\lfloor e^y\rfloor+1)=y+O(e^{-y})$. Thus the last term of $Q(y)$ contributes $(1+o(1))2y/\ln y$. This gives the asymptotic estimate
\begin{equation}\label{eq:ph-asymptotics}
Q(y)=(1+o(1))\frac{2y}{\ln y}\qquad\text{as }y\to\infty.
\end{equation}
We also need a bound for small guesses. Since $b(y)\ge2$, the formula for $Q(y)$ is bounded on every finite interval $[2,A]$. Together with~\eqref{eq:ph-asymptotics}, this gives a constant $C$ such that $Q(y)\le C(y+1)$ for every $y\ge2$.
We sum $Q(y_i)$ over the failed phases, precisely those with $y_i<\ln n$. Adding the bound from~\eqref{eq:ph-success} for the one successful phase and the cost of the final return gives
\begin{equation}\label{eq:ph-total}
\frac{\ALG}{\OPT}
\le \sum_{i:y_i<\ln n}Q(y_i)+2b^*+2+\frac{2\ln n}{\ln b^*}+\frac12.
\end{equation}
The last term pays for the final shortest path to $s$, by Lemma~\ref{lem:facts}(b).
Split the failed guesses at $Y=\ln n/(\ln\ln n)^2$. To bound the sum of the guesses below $Y$, observe that they grow at least geometrically. Indeed, $y_i=e^{h(t_i)}$, $t_{i+1}=t_i+1$, and $h'(t)=(1+\ln t)/2$ is increasing. Since $t_i\ge2$, the mean value theorem gives
\[
\frac{y_{i+1}}{y_i}=e^{h(t_i+1)-h(t_i)}\ge e^{h'(2)}=:\gamma>1.
\]
Reading backwards from the last guess below $Y$, the guesses are bounded by $Y,Y/\gamma,Y/\gamma^2,\ldots$. Also $y_i\ge2$, so $Q(y_i)\le C(y_i+1)\le 2Cy_i$. Thus all the small guesses together contribute at most
\[
\sum_{i:y_i<Y} Q(y_i)\le 2CY\sum_{j\ge0}\gamma^{-j}=O(Y)=o(\Lambda).
\]
For the remaining failed guesses, $Y\le y_i<\ln n$. Taking logarithms gives
\[
\ln\ln n-2\ln\ln\ln n=\ln Y\le\ln y_i<\ln\ln n.
\]
Both endpoints divided by $\ln\ln n$ tend to $1$. Hence $1/\ln y_i=(1+o(1))/\ln\ln n$ uniformly over this range. Since $Y\to\infty$, the relative error in~\eqref{eq:ph-asymptotics} also tends to zero uniformly for $y_i\ge Y$. We may therefore sum that estimate with a common error bound to obtain
\[
\sum_{i:Y\le y_i<\ln n} Q(y_i)
\le (1+o(1))\frac{2}{\ln\ln n}\sum_{i:Y\le y_i<\ln n}y_i.
\]
Adding the $o(\Lambda)$ contribution from guesses below $Y$ accounts for every failed phase. Extending the sum on the right to all failed guesses only increases it, since every $y_i$ is positive. This gives, uniformly in $\nu$, for the cost of all failed guesses
\begin{equation}\label{eq:ph-failed}
\sum_{i:y_i<\ln n}Q(y_i)
\le(1+o(1))\frac{2}{\ln\ln n}\sum_{i:y_i<\ln n}y_i+o(\Lambda).
\end{equation}

For the deterministic algorithm, set $\nu=0$. Let $j=i^*-1$ be the last failed phase index. The sum of logarithmic guesses on the right of~\eqref{eq:ph-failed} is $\sum_{i:y_i<\ln n}y_i=\sum_{i=0}^{j}y_i$. It suffices to bound this sum by $(1+o(1))\ln n$. As $n\to\infty$, we have $j\to\infty$. Since consecutive guesses grow by at least the factor $\gamma>1$ above, all guesses before $y_j$ sum to at most
\[
y_{j-1}(1+\gamma^{-1}+\gamma^{-2}+\cdots)=\frac{\gamma}{\gamma-1}y_{j-1}.
\]
This is $o(y_j)$ because $t_{j-1}=t_j-1$ and $y_i=t_i^{t_i/2}$, so $y_{j-1}\le t_j^{(t_j-1)/2}=y_j/\sqrt{t_j}$. Hence $\sum_{i=0}^j y_i\le(1+o(1))y_j\le(1+o(1))\ln n$. Substitution in~\eqref{eq:ph-failed} gives
\[
\sum_{i=0}^j Q(y_i)\le(1+o(1))\frac{2\ln n}{\ln\ln n}+o(\Lambda)
=(1+o(1))\Lambda.
\]
Combining this bound with~\eqref{eq:ph-success} in~\eqref{eq:ph-total} gives
\[
\frac{\ALG}{\OPT}\le
\underbrace{(1+o(1))\Lambda}_{\mathclap{\text{failed phases}}}
\quad+\quad\underbrace{(1+o(1))\Lambda}_{\mathclap{\text{successful phase}}}
\quad+\quad\underbrace{\tfrac12}_{\mathclap{\text{return to }s}}\,.
\]
The last term is $\Lambda\cdot\frac{\ln\ln n}{4\ln n}=o(\Lambda)$, since $\Lambda=2\ln n/\ln\ln n$. Thus the right-hand side is $(2+o(1))\Lambda=(4+o(1))\ln n/\ln\ln n$, proving (b).

For the randomized algorithm, phase $i$ contributes $y_i=e^{h(t_i)}$ to the sum in~\eqref{eq:ph-failed} exactly when $t_i<\hat{t}$. The expectation of its contribution is the integral over $\nu\in[0,1)$, since $\nu$ is uniform. Only indices with $i+2<\hat{t}$ can contribute, so linearity of expectation gives the first equality below.
\begin{equation}\label{eq:ph-average}
\E_\nu \Biggl[ \sum_{i:y_i<\ln n}y_i \Biggr]
=\sum_{i\ge0}\int_0^1
 \mathbf{1}_{\{i+2+\nu<\hat{t}\}}e^{h(i+2+\nu)}\,\mathrm{d}\nu
=\int_2^{\hat{t}} e^{h(z)}\,\mathrm{d}z.
\end{equation}
For the second equality, substitute $z=i+2+\nu$ in the $i$th integral. Then $\mathrm{d}z=\mathrm{d}\nu$ and $z$ ranges over $[i+2,i+3)$. These adjacent intervals cover $[2,\infty)$ exactly once, and the indicator retains precisely the portion $[2,\hat{t})$.

To show that this integral is $o(\ln n)$, recall that $e^{h(\hat{t})}=\ln n$ and $\hat{t}\to\infty$. With a real upper limit $x\to\infty$, both $\smash{\int_2^x e^{h(z)}\,\mathrm{d}z}$ and $\smash{e^{h(x)}}$ tend to infinity. The fundamental theorem of calculus gives the derivative $e^{h(x)}$ for the numerator, and the chain rule gives $h'(x)e^{h(x)}$ for the denominator. Hence l'Hôpital's rule yields
\[
\lim_{x\to\infty}\frac{\int_2^x e^{h(z)}\,\mathrm{d}z}{e^{h(x)}}
=\lim_{x\to\infty}\frac{1}{h'(x)}
=\lim_{x\to\infty}\frac{2}{1+\ln x}=0.
\]
Taking $x=\hat{t}$ in this estimate shows that the expected sum in~\eqref{eq:ph-average} is $o(\ln n)$. Since the errors in~\eqref{eq:ph-failed} are uniform in $\nu$, taking expectations there bounds the expected cost of the failed phases by $o(\Lambda)\OPT$.
Together with~\eqref{eq:ph-success} and~\eqref{eq:ph-total}, this proves (a).

Neither schedule uses $n$. For each fixed $n$, the number of phases and their bases are bounded uniformly in $\nu$, so~\eqref{eq:ph-total} gives a finite competitive bound. Hence the competitive ratio is finite for every $n$, and together with (a) and (b) it is $O(\log n/\log\log n)$ for all $n\ge2$.\qed
\end{proof}
\section{Tightness of the analysis}\label{sec:tight}

In this section, we show that our analysis for known $n$ is asymptotically tight and prove Theorem~\ref{thm:lower}.
In Section~\ref{app:tight}, we show that for every choice of $b=b(n)\ge2$, $\hDFSB$ has a competitive ratio of $\Omega(\log n/\log\log n)$. The bound is the larger of the bounds given by two constructions.
In Section~\ref{app:random-tight}, a cycle with an attached chain shows that for the base chosen in Theorems~\ref{thm:main} and~\ref{thm:offset}, the leading constant~$2$ is attained with and without the random offset. We prove no lower bound for $\hDFSBu$ with other bases.

\subsection{Lower bound for a fixed base}\label{app:tight}
Fix $n$ and the base $b\ge2$ used by $\hDFSB$. The first construction forces the agent to visit $n-2$ vertices along edges of weight $b$, although each is also reachable by an edge of weight just above $1$ in the same class. In the second, every new class starts another tree tour of the visited part of a path, and every such tour traverses a long segment that an optimal tour traverses only twice. We choose both instances to have exactly $n$ vertices. Theorem~\ref{thm:tight} takes the larger of their two lower bounds, so the result also covers a base chosen as a function of $n$.

\begin{figure}[tb]
\centering
\begin{tikzpicture}[font=\small]
  \coordinate (start) at (0,0);
  \coordinate (hub) at (5.6,0);
  \foreach \i/\height in {1/2.25,2/1.45,3/0.65} {
    \node[draw,circle,inner sep=1.5pt] (z\i) at (2.8,\height) {$z_{\i}$};
    \draw[thick] (start) -- (z\i);
    \draw[dashed] (hub) -- (z\i);
  }
  \path (start) -- (z1) node[midway,above left] {$b$};
  \path (z1) -- (hub) node[midway,above right] {$1+\eps$};
  \draw (start) -- node[midway,below] {$1$} (hub);
  \fill (start) circle (2pt) node[left=2pt] {$u_0=s$};
  \fill (hub) circle (2pt) node[right=2pt] {$u_1$};
\end{tikzpicture}
\caption{The trap for $p=3$. At level $1$, the agent reaches each $z_i$ from $u_0$ along a thick edge of weight $b$. A tour can visit the same vertices from $u_1$ along the dashed edges of weight $1+\eps$.}
\label{fig:trap}
\end{figure}
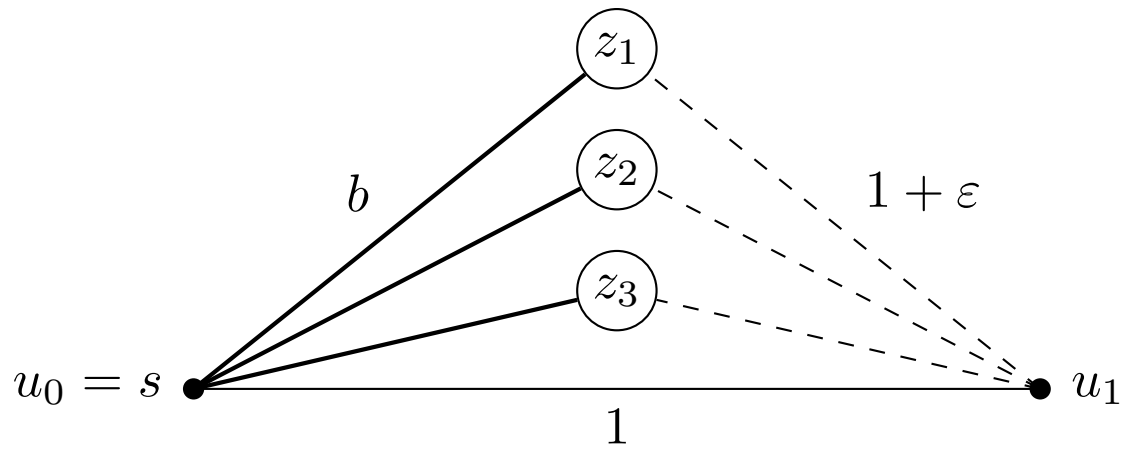
\begin{lemma}\label{lem:trap}
For $b\ge2$, an integer $p\ge1$, and $0<\eps\le b-1$, there is an instance with $p+2$ vertices satisfying
$\ALG(\hDFSB)\ge 2 + 2pb$ and $\OPT\le2+2p(1+\eps)$.
\end{lemma}
\begin{proof}
Consider the \emph{trap} in Fig.~\ref{fig:trap} on the vertices $u_0=s,u_1,z_1,\ldots,z_p$. Join $u_0$ to $u_1$ by an edge of weight~$1$. For each $i=1,\ldots,p$, add the edges $\{u_0,z_i\}$ and $\{u_1,z_i\}$ of weights $b$ and $1+\eps$, respectively.
The edge of weight $1$ has class $0$. Since $0<\eps\le b-1$, both $1+\eps$ and $b$ lie in $(1,b]$, the interval defining class $1$.

The level-$0$ iteration visits $u_1$ and returns, at cost $2$.
At level $1$, the cluster $\{u_0,u_1\}$ is a single quotient vertex, and each $z_i$ is a singleton neighbour. The tree tour starts at $u_0$, where \Process{} examines every edge $\{u_0,z_i\}$ before moving to $u_1$. Each forward traversal and its return cost $2b$, giving the claimed lower bound $2+2pb$.
The tour that visits every $z_i$ from $u_1$ costs $2+2p(1+\eps)$.\qed
\end{proof}

The next construction adapts the weighted path used by Megow et al.~\cite[proof of Theorem~5]{megow2012} to prove their lower bound for power-of-two $\hDFS$.

\begin{lemma}\label{lem:path}
For $b\ge2$, an integer $\ell\ge1$, and an even integer $D\ge2$, there is a path with $D+\ell+1$ vertices such that
$\ALG(\hDFSB)\ge2(\ell+1)D+2H$, $\OPT=2(D+H)$, $H=\sum_{j=1}^{\ell}b^j$.
If $D\ge(\ell+1)b^{\ell+1}$, then $\ALG(\hDFSB)/\OPT\ge \ell$ on this path.
\end{lemma}
\begin{proof}
Let $s$ be the centre of a path of $D$ unit edges with endpoints $a$ and $c$ (Fig.~\ref{fig:path}).
Attach edges $h_1,\ldots,h_\ell$ of weights $b,b^2,\ldots,b^\ell$, alternately extending the right and left ends. Thus each of the two attached chains has increasing weights outwards. Let $x_j$ be the outer endpoint of $h_j$.

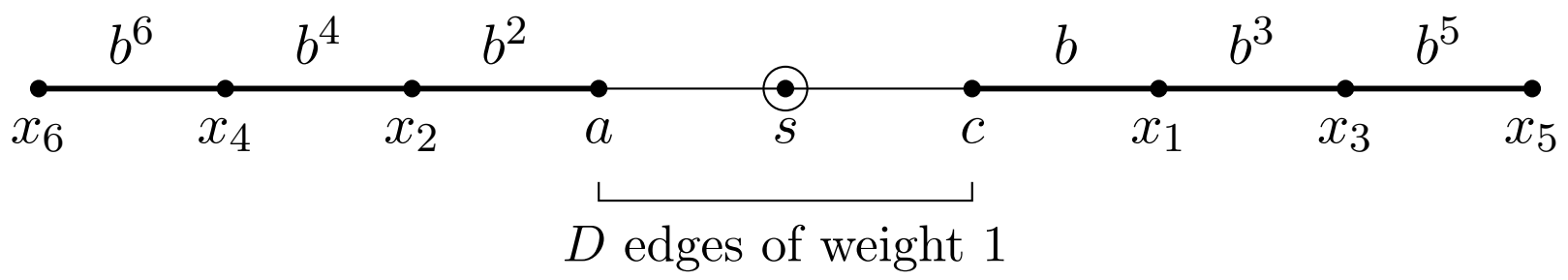
\begin{figure}[tb]
\centering
\begin{tikzpicture}[scale=0.85]
  \draw[line width=1.1pt] (0,0) -- (4.5,0);
  \draw[line width=0.4pt] (4.5,0) -- (7.5,0);
  \draw[line width=1.1pt] (7.5,0) -- (12,0);
  \foreach \x/\nm in {0/x_6,1.5/x_4,3/x_2,4.5/a,6/s,7.5/c,9/x_1,10.5/x_3,12/x_5} {
    \fill (\x,0) circle (2pt);
    \node[below=2pt] at (\x,0) {$\nm$};
  }
  \draw (6,0) circle (5pt);
  \foreach \x/\wt in {0.75/b^6,2.25/b^4,3.75/b^2,8.25/b,9.75/b^3,11.25/b^5} {
    \node[above=1pt] at (\x,0) {$\wt$};
  }
  \draw[line width=0.4pt] (4.5,-0.75) -- (4.5,-0.9) -- (7.5,-0.9) -- (7.5,-0.75);
  \node[below=1pt] at (6,-0.9) {\small $D$ edges of weight $1$};
\end{tikzpicture}
\caption{The path of Lemma~\ref{lem:path} for $\ell=6$. Each heavy edge starts a new level, whose tree tour traverses the whole unit segment twice.}
\label{fig:path}
\end{figure}

The unit edges have class $0$ and $h_j$ has class $j$. Hence, for $0\le j\le \ell$, the level-$j$ cluster $X_j(s)$ consists of the unit segment and the vertices $x_1,\ldots,x_j$.

We first bound the cost of the top-level iterations. By Lemma~\ref{lem:invariant}(c), iteration $0$ starts at $s$ with $S=\{s\}$ and ends at $s$ with $S=X_0(s)$, and by Lemma~\ref{lem:invariant}(a) its walk stays in the unit segment. A closed walk in a path that visits both endpoints traverses every edge at least twice, so iteration $0$ costs at least $2D$.
For $j=1,\ldots,\ell$, the vertex $x_j$ is still unvisited after iteration $j-1$, so the main loop performs iteration $j$. It starts at $s$ with $S=X_{j-1}(s)$, again by Lemma~\ref{lem:invariant}(c). The edge $h_j$ is a class-$j$ boundary edge leaving this cluster, so $\Process{}(j,X_{j-1}(s),s)$ starts a tree tour of $X_{j-1}(s)$ and runs it to completion. The tree of this cluster is the subpath on $X_{j-1}(s)$, which contains all $D$ unit edges, and the tour traverses each of its edges twice. Together with the two traversals of $h_j$, iteration $j$ costs at least $2D+2b^j$. Summing over all iterations gives
\[
\ALG(\hDFSB)\ge2D+\sum_{j=1}^\ell\bigl(2D+2b^j\bigr)=2(\ell+1)D+2H.
\]
Since the graph is a tree, doubling all its edges is optimal, giving $\OPT=2(D+H)$.

It remains to bound the ratio. Since $b\ge2$, we have $H=(b^{\ell+1}-b)/(b-1)\le2b^\ell$. Together with the assumption $D\ge(\ell+1)b^{\ell+1}$, this gives $(\ell+1)H/D\le2/b\le1$. Hence
\[
\frac{\ALG(\hDFSB)}{\OPT}
\ge\frac{(\ell+1)D}{D+H}
=\frac{\ell+1}{1+H/D}
\ge(\ell+1)\Bigl(1-\frac HD\Bigr)
\ge \ell+1-\frac2b
\ge \ell.\tag*{\qed}
\]
\end{proof}

For a fixed number $n$ of vertices, the trap in Lemma~\ref{lem:trap} gives a stronger lower bound as $b$ increases, while the path in Lemma~\ref{lem:path} forces more tours of its central segment for smaller $b$. Taking the larger of the two bounds gives a lower bound that holds even when the base depends on $n$.

\begin{theorem}\label{thm:tight}
Let $n\ge16$ and $b\ge2$.
The competitive ratio of $\hDFSB$ on instances with exactly $n$ vertices is at least
\[
\max\left\{b\Bigl(1-\frac{2}{n}\Bigr),\,
\frac{\log_2n-\log_2\log_2n-1}{\log_2b}-3\right\}.
\]
This maximum is at least $\log_2 n/\log_2\log_2 n-5$. Thus $\hDFSB$ has a competitive ratio of $\Omega(\log n/\log\log n)$ for every choice $b=b(n)\ge2$.
\end{theorem}
\begin{proof}
The trap with $p=n-2$ and $\eps=1/n\le b-1$ has exactly $n$ vertices and $\OPT<2n$, so $\ALG(\hDFSB)/\OPT>b(1-2/n)$ on this instance.
For the second bound put
\[
x=\log_b\frac{n}{2\log_2n}
=\frac{\log_2n-\log_2\log_2n-1}{\log_2b}.
\]
If $x<4$, the claimed bound $x-3$ is less than $1$. It needs no construction, since the walk of $\hDFSB$ is a closed walk from $s$ that visits every vertex (Lemma~\ref{lem:invariant}(c)), so $\ALG(\hDFSB)\ge\OPT$ on every instance.
Otherwise choose the largest integer $\ell+1\le x$ having the parity of $n$, and let $D=n-\ell-1$. Then $\ell+1>x-2\ge2$, $D$ is even, and
\[
(\ell+1)b^{\ell+1}\le\log_2n\cdot\frac{n}{2\log_2n}
=\frac n2\le D,
\]
since $b^{\ell+1}\le b^x=n/(2\log_2n)$ and $\ell+1\le x<\log_2n\le n/2$ for $n\ge16$ (recall $\log_2b\ge1$).
Lemma~\ref{lem:path} gives an $n$-vertex instance with $\ALG(\hDFSB)/\OPT\ge \ell>x-3$.

Write $L=\log_2n/\log_2\log_2n$. If $b\ge L$, the trap gives at least $L(1-2/n)\ge L-1/2$, since $2L/n\le2\log_2n/n\le1/2$ for $n\ge16$. If $b<L$, then $1\le\log_2b\le\log_2\log_2n$. The numerator $\log_2n-\log_2\log_2n-1$ is positive for $n\ge16$, so the second bound is at least
\[
L-1-\frac{1}{\log_2\log_2n}-3
\ge L-\frac92>L-5.
\]
Both constructions have exactly $n$ vertices, so the argument also applies when $b$ depends on $n$.\qed
\end{proof}
\subsection{Tightness of the leading constant}\label{app:random-tight}

To force repeated travel inside a cluster, attach a chain to a long cycle. We choose each chain weight to be an integer power of the prescribed base $b$. The random offset then preserves the class of every edge.

\begin{theorem}\label{thm:random-tight}
Let $b=\max\{2,\ln n/(\ln\ln n)^2\}$. For every sufficiently large $n$, there is an $n$-vertex instance on which
\[
\ALG(\hDFSBu)\ge(2-o(1))\frac{\ln n}{\ln\ln n}\OPT\qquad\text{for every }u\in[0,1).
\]
In particular, the bound holds for $\hDFSB$, which is the case $u=0$, and for $\E_u[\ALG(\hDFSBu)]$ with $u$ drawn uniformly from $[0,1)$.
\end{theorem}
\begin{proof}
Set $\ell=\lfloor\log_b(n/\ln n)\rfloor$ and $D=n-\ell$. Start at a vertex $s$ of a cycle of $D$ unit edges and attach at $s$ a chain with edge weights $b,b^2,\ldots,b^\ell$ in order away from $s$ (Fig.~\ref{fig:cycle-chain}). For every $u\in[0,1)$ and integer $j$, the class of an edge of weight $b^j$ is $\lceil j-u\rceil=j$. Thus the algorithm visits the cycle at level $0$ and reaches one new chain vertex at each level $j=1,\ldots,\ell$.

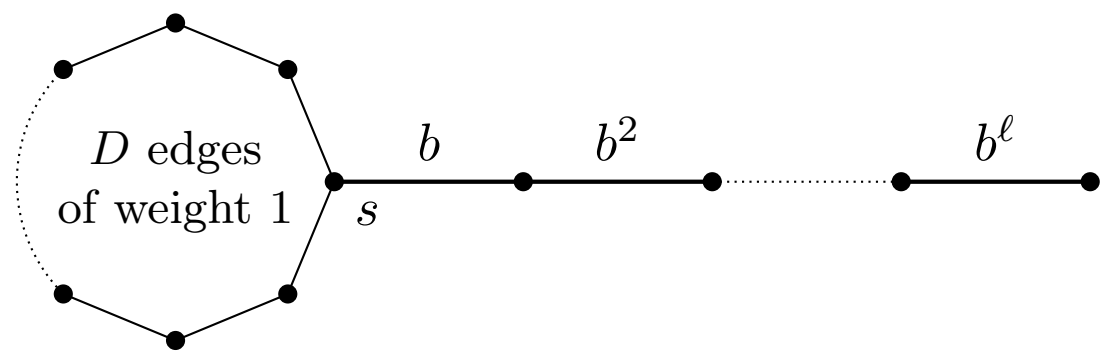
\begin{figure}[htb]
\centering
\begin{tikzpicture}[font=\small]
  \foreach \angle in {0,45,90,135,225,270,315}
    \coordinate (v\angle) at (\angle:1.05);
  \draw (v225) -- (v270) -- (v315) -- (v0) -- (v45) -- (v90) -- (v135);
  \draw[densely dotted] (v135) arc[start angle=135,end angle=225,radius=1.05];
  \node[align=center,font=\footnotesize] at (0,0) {$D$ edges\\of weight $1$};
  \draw[thick] (v0) -- node[above] {$b$} (2.3,0)
                       -- node[above] {$b^2$} (3.55,0);
  \draw[densely dotted] (3.55,0) -- (4.8,0);
  \draw[thick] (4.8,0) -- node[above] {$b^\ell$} (6.05,0);
  \foreach \angle in {0,45,90,135,225,270,315}
    \fill (v\angle) circle (1.8pt);
  \foreach \x in {2.3,3.55,4.8,6.05}
    \fill (\x,0) circle (1.8pt);
  \node[below right] at (v0) {$s$};
\end{tikzpicture}
\caption{The instance of Theorem~\ref{thm:random-tight}. Each successive chain edge starts a new level, forcing another tour of a spanning tree of the visited cluster.}
\label{fig:cycle-chain}
\end{figure}

At the start of level $j$, the cluster containing $s$ consists of the cycle and the first $j-1$ chain edges. The next chain edge is a class-$j$ boundary edge, so \Process{} starts and completes a tree tour of this cluster. Every spanning tree of the cluster contains $D-1$ cycle edges. The tour crosses each of them twice, giving a cost of at least $2(D-1)\ell$ for every offset.

An offline tour goes around the cycle once and traverses the chain in both directions. The chain has total weight at most $2b^\ell\le 2n/\ln n$, since $b\ge2$. Thus $\OPT\le D+4n/\ln n$. For the stated base, $\ln b=\ln\ln n-2\ln\ln\ln n$ for large $n$, so
\[
\ell=\left\lfloor\frac{\ln n-\ln\ln n}{\ln b}\right\rfloor
=(1+o(1))\frac{\ln n}{\ln\ln n}.
\]
In particular $\ell=o(n)$ and $D=(1-o(1))n$. Dividing the cost bound by $\OPT$ gives at least $2\ell(D-1)/(D+4n/\ln n)=(2-o(1))\ln n/\ln\ln n$. Neither the cost bound nor the bound on $\OPT$ depends on $u$, so the ratio bound holds for every $u\in[0,1)$, in particular for $u=0$ and after averaging over $u$.\qed
\end{proof}
\end{document}